\documentclass[conference]{IEEEtran}
\IEEEoverridecommandlockouts

\usepackage{cite}
\usepackage{amsmath,amssymb,amsfonts}
\usepackage{graphicx}
\usepackage{textcomp}
\usepackage{multirow}
\usepackage{xcolor}
\usepackage{booktabs}
\usepackage[ruled,vlined]{algorithm2e}
\usepackage{url,hyperref,comment}
\usepackage{enumitem}

\newcommand{\TA}{T-Audio (DP)}
\newcommand{\TMM}{T-MM (DP; audio-only eval)}
\newcommand{\SA}{S-KD(\TA)}
\newcommand{\SMMa}{S-KD(\TMM; query=audio)}
\newcommand{\SMMp}{S-KD(\TMM; query=priv)}

\def\BibTeX{{\rm B\kern-.05em{\sc i\kern-.025em b}\kern-.08em
    T\kern-.1667em\lower.7ex\hbox{E}\kern-.125emX}}

\begin{document}

\title{Private Speech Classification without Collapse: Stabilized DP Training and Offline Distillation}

\author{
  Yadi Wen$\dagger^1$, Tianxin Li$\dagger^2$, Enji Liang$^1$, Rong Du$^1*$, and Yue Fu$^{1}$
  \thanks{This paper has been accepted for publication at the ICME 2026 Workshop. * Corresponding author, $\dagger$ equal contributor.}
  \\
  \IEEEauthorblockA{
    1. Shanxi Key Laboratory of Industrial Internet Security, Taiyuan University of Technology, China\\
    2. University of Michigan, Ann Arbor, USA\\
    Emails: 2023005322@link.tyut.edu.cn, ltianxin@umich.edu, liangenji@tyut.edu.cn, durong@tyut.edu.cn, fuyue@tyut.edu.cn
  }
}
\maketitle

\begin{abstract}
We study example-level private supervised speech classification under a practical release constraint: training may access privileged side information, but the released model must be audio-only. This setting is important because speech systems can often exploit richer side information during development, whereas deployment and release require a lightweight unimodal model with auditable privacy guarantees. Using DP-SGD on the private dataset $D_{\text{priv}}$, we identify a strong-privacy failure mode ($\epsilon \le 1$) on imbalanced tasks, where training may collapse to a near single-class predictor, a phenomenon that overall accuracy can obscure. We therefore emphasize Macro-F1, balanced accuracy, and a simple collapse diagnostic. This failure is especially problematic in our release setting because a collapsed private teacher cannot provide useful supervision for the downstream audio-only student.

To address this setting under strong privacy, we propose a two-stage protocol: (i) train a (possibly multimodal) DP teacher on $D_{\text{priv}}$, and (ii) distill an audio-only student on a fixed, recording-disjoint auxiliary dataset $D_{\text{aux}}$ using one-shot offline teacher probability outputs, releasing only the student. The DP guarantee applies only to $D_{\text{priv}}$; we make no DP claim for $D_{\text{aux}}$, and privacy of the released student with respect to $D_{\text{priv}}$ follows by post-processing. We frame this setting as involving four coupled bottlenecks: speech-induced optimization instability under DP-SGD, minority-class erosion under clipping and noise, teacher over-reliance on privileged modalities unavailable at deployment, and train--deploy modality mismatch. We address them with a DP-stabilizing acoustic front-end (DSAF), minibatch-adaptive bounded loss reweighting (AW-DP), privileged-modality dropout, and offline teacher-to-student distillation.
\end{abstract}


\section{Introduction}
\label{sec:intro}

Speech models underpin many multimedia systems, including spoken-content indexing, recommendation, and voice interaction. Their training data are highly sensitive: user utterances reveal not only linguistic content but also speaker identity and incidental context. Meanwhile, trained models are often released as checkpoints or exposed through prediction APIs, making parameters or confidence scores accessible to potentially untrusted parties. Prior work has shown that such artifacts can leak training-set information via membership inference and model inversion attacks~\cite{shokri2017membershipInference,fredrikson2015modelInversion}. These risks motivate training and release procedures with explicit, auditable privacy guarantees. Differential Privacy (DP) bounds the influence of any single example on a released artifact~\cite{dwork2006DP,dwork2008surveyDP,dwork2014}. In deep learning, example-level DP-SGD enforces this guarantee by clipping per-sample gradients and adding Gaussian noise to the aggregated update~\cite{abadi2016deepLearningDP}. However, speech models typically operate on log-Mel time-frequency features~\cite{OShaughnessy1987SpeechCommunication}, where variation in utterance length and energy induces substantial gradient-norm heterogeneity, increasing clipping and degrading effective signal-to-noise~\cite{chen2020understandingClipping}. Class imbalance further compounds the problem: under private minibatch sampling, minority classes may contribute weak and noisy updates, leading to systematic underfitting~\cite{tran2021dpimbalance}. Although adaptive clipping and related scheduling strategies improve robustness~\cite{xia2023perSampleAdaptiveClipping,chilukoti2025dpsgdglobadaptv2s}, strong-privacy training remains especially difficult in speech. We view this as a release-constrained learning problem in which privacy, imbalance, and speech-specific optimization instability interact.

A central failure mode in this setting is prediction collapse. On imbalanced speech classification, we observe that DP-SGD under strong privacy (e.g., $\epsilon \le 1$ in our setup) can converge to a near single-class, typically majority-class, predictor. Because overall accuracy may remain deceptively high under imbalance, we do not treat it as a primary utility metric. Instead, we evaluate utility using Macro-F1 and balanced accuracy (Bal-Acc), the mean per-class recall, and quantify collapse with Maj-Pred, the fraction of predictions assigned to the most frequently predicted class. Maj-Pred is used only as a diagnostic: for a $K$-class task, a single-class predictor yields $\mathrm{Bal\text{-}Acc}=1/K$ and $\mathrm{Maj\text{-}Pred}=1$ regardless of class proportions. This failure is consequential not only because it degrades the private model, but also because a collapsed teacher cannot provide informative soft targets for transfer.

A second challenge is modality mismatch. In many multimedia applications, training can exploit text or metadata, whereas deployment must be audio-only. Paired audio-text or audio-metadata resources are common~\cite{ardila2020commonvoice}, and multimodal learning often improves robustness~\cite{baltrusaitis2019multimodalSurvey}. Under privacy constraints, however, a multimodal teacher cannot simply be exposed after training; any useful information learned from private data must be transferred into a deployable audio-only artifact. This setting naturally connects to Learning Using Privileged Information (LUPI)~\cite{vapnik2009lupi} and knowledge distillation~\cite{hinton2015distilling,lopezpaz2016unifying}. We therefore adopt the following release protocol: train a (possibly multimodal) DP teacher on private data $D_{\text{priv}}$; obtain teacher probability outputs on a recording-disjoint auxiliary set $D_{\text{aux}}$; distill an audio-only student on $D_{\text{aux}}$; and release only the student. Because the teacher is DP with respect to $D_{\text{priv}}$, any artifact computed solely from its outputs inherits the same privacy guarantee by post-processing~\cite{dwork2014}. Using a fixed, predetermined $D_{\text{aux}}$ also avoids adaptive-querying concerns.

Our setting differs from three closely related lines of work. First, unlike prior DP-SGD stabilization methods, we study privacy-preserving speech learning under a release constraint: utility must survive transfer to a deployable audio-only model rather than be measured only on the private model. Second, unlike standard distillation or LUPI, the teacher is DP-trained on $D_{\text{priv}}$ and never released. Third, unlike interactive private prediction settings, we use a one-shot offline labeling protocol on fixed auxiliary data, yielding a simple and auditable release mechanism.

The resulting objective is not merely to improve private teacher accuracy, but to maintain stable private optimization, preserve minority-class structure, avoid over-reliance on privileged modalities, and produce soft labels suitable for offline transfer. To this end, we propose a coordinated pipeline: DSAF reduces speech-induced gradient-norm drift before clipping; AW-DP mitigates imbalance-amplified minority erosion during DP training; privileged-modality dropout discourages shortcut reliance on training-only side information; and offline distillation converts the DP teacher into the only deployable artifact, an audio-only released model. The contributions of this paper are three-fold:

\begin{figure}[]
  \centering
  \includegraphics[width=\linewidth]{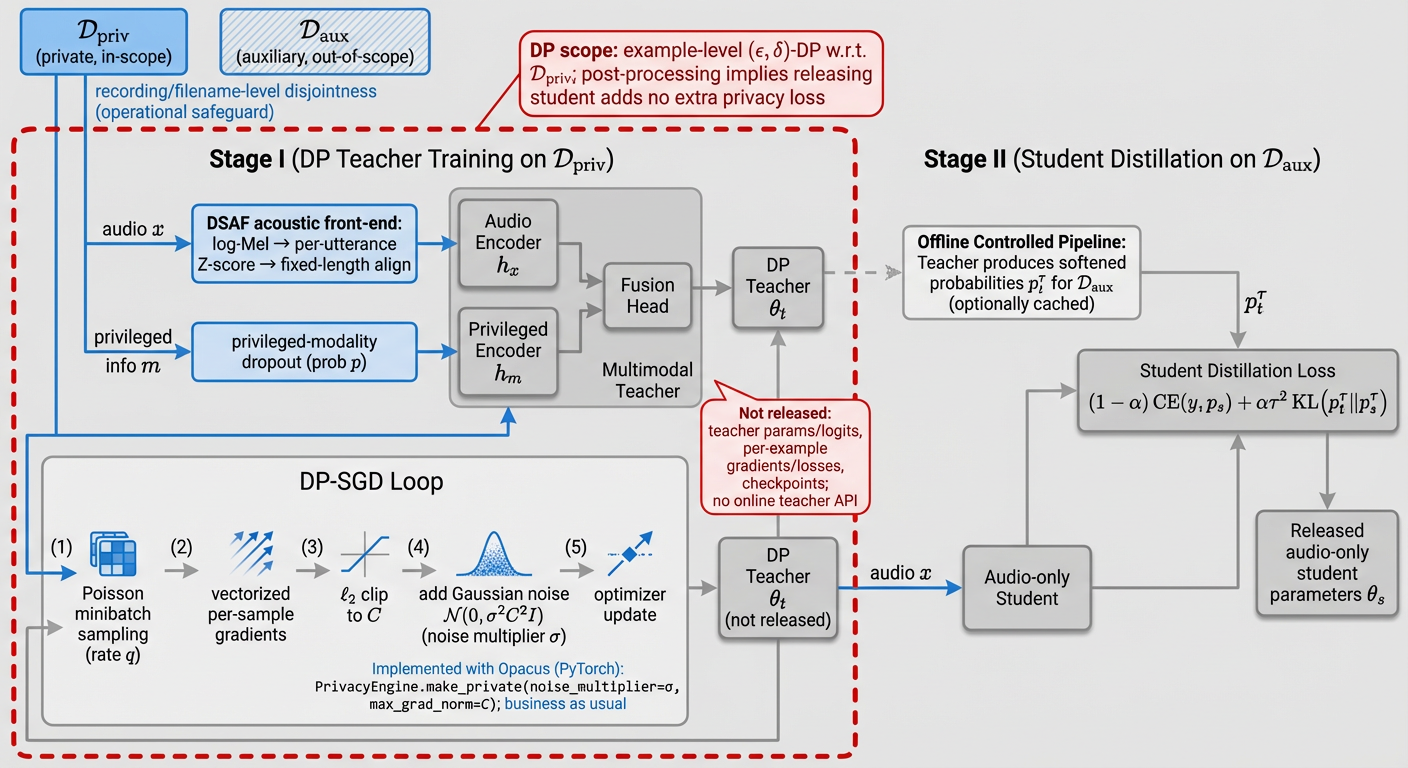}
  \caption{Workflow of the proposed method.}
  \vspace{-5mm}
  \label{fig:pipeline}
\end{figure}

\begin{itemize}
    \item \textbf{Problem framing and failure mode.}
    We formulate private speech classification under a release constraint where training may use privileged side information but the released model must be audio-only. In this setting, we identify a strong-privacy failure mode of DP-SGD on imbalanced speech tasks: collapse to a near single-class predictor. Accordingly, we emphasize Macro-F1, balanced accuracy, and Maj-Pred rather than overall accuracy alone.

    \item \textbf{DP release protocol under modality mismatch.}
    We propose a two-stage protocol that trains a (possibly multimodal) DP teacher on $D_{\text{priv}}$, labels a fixed auxiliary set $D_{\text{aux}}$ offline, distills an audio-only student, and releases only the student. The privacy guarantee with respect to $D_{\text{priv}}$ follows directly by post-processing.

    \item \textbf{Teacher stabilization for transferable private supervision.}
    We introduce DSAF, AW-DP, and privileged-modality dropout to improve the stability, class balance, and audio-transferability of DP teacher predictions under strong privacy, thereby improving the released student obtained by offline distillation.
\end{itemize}

\section{Preliminaries}
\label{sec:prelim}

\subsection{Setup and release setting}
We study supervised speech classification with training examples
$\{(x_i,m_i,y_i)\}_{i=1}^N$, where $x_i$ is audio, $m_i$ is privileged side information available only during training (e.g., transcript or metadata), and $y_i\in\{1,\dots,K\}$ with $K{=}3$ for \texttt{male/female/other}. At inference, only $x$ is available, so the released model must be audio-only. We use this 3-way label space as an imbalanced benchmark for analyzing DP-SGD failure modes under modality mismatch. This choice is purely methodological and is not intended to advocate deployment of gender inference systems.

We distinguish two datasets. $D_{\text{priv}}$ is the private in-scope dataset used to train a differentially private teacher. $D_{\text{aux}}$ is an auxiliary out-of-scope dataset used to train the released student via distillation. We enforce recording/filename-level disjointness between $D_{\text{priv}}$ and $D_{\text{aux}}$ as an operational safeguard against overlap-induced leakage and to better reflect deployment transfer; this disjointness is not required for the DP guarantee, which is defined only with respect to $D_{\text{priv}}$.

\subsection{Threat model and privacy scope}
The adversary obtains the released student parameters $\theta_s$ (white-box access), which subsumes black-box access to the released audio-only model $f_s(\cdot)$. The privacy scope is $D_{\text{priv}}$: our goal is example-level $(\epsilon,\delta)$-DP with respect to the inclusion or removal of any single record in $D_{\text{priv}}$. We release only the student; the teacher is not released. We make no DP claim for $D_{\text{aux}}$, since the student is trained non-privately on $D_{\text{aux}}$; accordingly, we assume $D_{\text{aux}}$ is public or consented for such use.

\subsection{Example-level differential privacy}
A randomized algorithm $\mathcal{A}$ is $(\epsilon,\delta)$-differentially private (DP) if for any neighboring datasets $D,D'$ differing in one example and any measurable set $S$,
\begin{equation}
\Pr[\mathcal{A}(D)\in S] \le e^\epsilon \Pr[\mathcal{A}(D')\in S] + \delta.
\end{equation}

We use the post-processing property: if $\mathcal{M}$ is $(\epsilon,\delta)$-DP, then for any (possibly randomized) transformation $\mathcal{T}$, $\mathcal{T}\!\circ\!\mathcal{M}$ is also $(\epsilon,\delta)$-DP~\cite{dwork2014}. In our pipeline, teacher training defines $\theta_t \leftarrow \mathcal{M}(D_{\text{priv}})$ and student training defines $\theta_s \leftarrow \mathcal{T}(D_{\text{aux}},\theta_t)$; therefore releasing $\theta_s$ incurs no additional privacy loss with respect to $D_{\text{priv}}$.

\subsection{DP-SGD and privacy accounting}
We instantiate example-level DP using DP-SGD with per-sample $\ell_2$ clipping and Gaussian noise~\cite{abadi2016deepLearningDP,dwork2014}. Let $g_i=\nabla_{\theta}\ell(\theta; x_i,m_i,y_i)$ and
$\mathrm{clip}(g_i,C)=g_i\cdot \min\{1, C/\lVert g_i\rVert_2\}$.
With Poisson subsampling rate $q$ and sampled minibatch $\mathcal{B}$, the privatized update direction is
\begin{equation}
\tilde{g}
=
\frac{1}{|\mathcal{B}|}
\left(
\sum_{i\in\mathcal{B}} \mathrm{clip}(g_i,C)
+
\mathcal{N}(0,\sigma^2 C^2 I)
\right),
\end{equation}
where $C$ is the clipping threshold and $\sigma$ is the noise multiplier.

We report composed privacy budgets $(\epsilon,\delta)$ using an RDP accountant~\cite{mironov2017rdp} under the exact training configuration (sampling rate $q$, number of steps, $C$, $\sigma$, and $\delta$). To support per-sample gradient computation, our DP teacher models avoid minibatch-coupled state.

\subsection{Knowledge distillation}
We use standard knowledge distillation to transfer information from teacher to student~\cite{hinton2015distilling}. Let $p_t$ and $p_s$ denote teacher and student class probabilities. With temperature $\tau$ and mixing coefficient $\alpha$, the student objective on the auxiliary dataset is
\begin{equation}
\mathcal{L}_{\text{KD}}
=
(1-\alpha)\,\mathrm{CE}(y,p_s)
+
\alpha\,\tau^2\,\mathrm{KL}(p_t^\tau \| p_s^\tau),
\end{equation}
where $p^\tau$ denotes the probability vector under temperature $\tau$.

\section{Method}
\label{sec:method}

\subsection{Release protocol}
\label{sec:release_protocol}
We consider a release setting in which a teacher is trained on a private dataset $D_{\text{priv}}$, while the final released model must be an audio-only classifier. We adopt a two-stage protocol: (i) train a differentially private (DP) teacher on $D_{\text{priv}}$; (ii) query the teacher once on a fixed non-private auxiliary dataset $D_{\text{aux}}$ and train an audio-only student by offline distillation. Only the student is released. By DP post-processing, releasing $\theta_s$ incurs no privacy loss beyond that of the DP teacher.

\subsection{Overview}
\label{sec:method_overview}
Under strong privacy, a teacher may retain non-trivial accuracy yet still provide poor supervision for student release: its posterior can become over-smoothed, majority-dominated, overly dependent on privileged inputs, or poorly aligned with the deployment-time audio-only model. We therefore organize the method around four coupled bottlenecks in private speech learning: (1) unstable acoustic inputs under DP-SGD, (2) imbalance-amplified minority suppression under clipping and noise, (3) teacher over-reliance on privileged modalities unavailable at deployment, and (4) modality mismatch between a possibly multimodal private teacher and the released audio-only student.

Specifically, we combine:
(1) \textbf{DP-Stabilizing Acoustic Front-End (DSAF)} to reduce input-induced gradient-norm dispersion;
(2) \textbf{Imbalance-aware Weighted DP-SGD (AW-DP)} to preserve minority-class signal under clipping and noise;
(3) \textbf{privileged-modality dropout} to discourage shortcut reliance on privileged inputs and improve audio-transferable predictions; and
(4) \textbf{offline teacher-to-student distillation} to resolve the train--deploy modality mismatch while preserving privacy by post-processing.
Together, these components should be understood as a unified release pipeline rather than as independent add-ons.

\subsection{Stage I: DP teacher training}
\label{sec:method_teacher}
The teacher is trained on $D_{\text{priv}}$ with DP-SGD. Depending on available inputs, it can be either audio-only or multimodal with privileged side information. Unless disabled in ablations, DSAF and AW-DP are always applied; privileged-modality dropout is used only in the multimodal setting.

\subsubsection{Teacher architecture}
\label{sec:teacher_arch}
For the multimodal teacher, we use an audio encoder $h_x(\cdot)$, a privileged-information encoder $h_m(\cdot)$, and a fusion head $\phi(\cdot)$:
\begin{equation}
p_t(x,m)=\mathrm{softmax}\!\left(\phi([h_x(X);\;h_m(m')])\right),
\end{equation}
where $X$ is the processed acoustic feature and $m'$ is the possibly dropped privileged input. Privileged-modality dropout is defined as
\begin{equation}
m'=
\begin{cases}
0, & \text{with probability } p,\\
m, & \text{with probability } 1-p.
\end{cases}
\end{equation}
This discourages over-reliance on privileged cues and promotes audio-transferable teacher predictions.

For an audio-only teacher, we use
\begin{equation}
p_t(x)=\mathrm{softmax}\!\left(\phi(h_x(X))\right).
\end{equation}

\subsubsection{DP-SGD}
\label{sec:dpsgd_workflow}
Each training step uses Poisson minibatch sampling, per-sample gradient computation, global $\ell_2$ clipping with threshold $C$, Gaussian noise with multiplier $\sigma$, and parameter update. Given per-example gradients $g_i=\nabla_{\theta_t}\ell_i$, we compute
\begin{equation*}
\bar g_i = g_i \cdot \min\left\{1,\frac{C}{\|g_i\|_2}\right\},
\end{equation*}
and apply the noisy update
\begin{equation}
\tilde g
=
\frac{1}{|\mathcal{B}|}
\left(
\sum_{i\in\mathcal{B}} \bar g_i
+
\mathcal{N}(0,\sigma^2 C^2 I)
\right).
\end{equation}
Our method does not alter this privacy mechanism; instead, it improves the stability and usefulness of the signal entering DP-SGD.

\subsubsection{DP-Stabilizing Acoustic Front-End (DSAF)}
\label{sec:dsaf}
Speech inputs can yield highly variable per-sample gradient norms, increasing over-clipping under DP-SGD. To reduce this effect, we compute log-Mel spectrograms $S$ and apply per-utterance normalization with fixed-length alignment:
\begin{equation}
X=
\mathrm{FixLen}\!\left(
\frac{S-\mu(S)}{\sigma(S)+\eta_0},
L
\right),
\end{equation}
where $\mu(S)$ and $\sigma(S)$ are computed from the current utterance, $\eta_0>0$ is a stabilizer, and $\mathrm{FixLen}(\cdot,L)$ truncates or pads to $L$ frames. DSAF is deterministic and per-example, and is therefore fully compatible with DP-SGD.

\subsubsection{Imbalance-aware Weighted DP-SGD (AW-DP)}
\label{sec:aw_dpsgd}
To mitigate minority-class suppression under strong privacy, we reweight examples at the loss level before clipping. For minibatch $\mathcal{B}_t$ with class counts $n_k^{(t)}$, we define
\begin{equation}
w_k^{(t)}=\frac{|\mathcal{B}_t|}{K\cdot n_k^{(t)}+\varepsilon_w},
\end{equation}
with $\varepsilon_w>0$ and $w_k^{(t)}=1$ when $n_k^{(t)}=0$. We then clip the weights:
\begin{equation}
w_k^{(t)} \leftarrow \min\{w_{\max},\max\{w_{\min},w_k^{(t)}\}\}.
\end{equation}
The weighted loss is
\begin{equation}
\ell_i = w_{y_i}^{(t)}\,\mathrm{CE}\!\bigl(y_i,p_t(x_i,m_i)\bigr),
\qquad
\mathcal{L}=\frac{1}{|\mathcal{B}_t|}\sum_{i\in\mathcal{B}_t}\ell_i.
\end{equation}
For audio-only teachers, $p_t(x_i,m_i)$ reduces to $p_t(x_i)$. AW-DP changes only the objective, not the privacy mechanism; the same accountant applies under fixed $(q,T,\sigma,C,\delta)$.

\subsection{Stage II: Distillation to the released audio-only student}
\label{sec:method_student}
We train an audio-only student $p_s(x)$ on $D_{\text{aux}}$ using offline teacher predictions. This stage is the mechanism that resolves the train--deploy modality mismatch: the teacher may use privileged inputs during private training, but the only released artifact is an audio-only student. The teacher is queried once on a fixed $D_{\text{aux}}$ and is not released.

The effectiveness of this stage depends directly on the quality of the DP teacher outputs. If teacher predictions collapse toward majority classes or rely excessively on privileged information unavailable to the student at deployment, distillation yields limited gains. In contrast, DSAF, AW-DP, and privileged-modality dropout improve the stability, class balance, and transferability of teacher outputs, leading to a stronger released student.

Let $\tau$ denote the temperature and $\alpha$ the mixing coefficient. We optimize
\begin{equation}
\mathcal{L}_{\text{KD}}
=
(1-\alpha)\,\mathrm{CE}(y,p_s)
+
\alpha\,\tau^2\,\mathrm{KL}(p_t^\tau \| p_s^\tau),
\end{equation}
where $p^\tau$ denotes probabilities computed with temperature $\tau$.

\begin{algorithm}[]
\caption{DP-SGD training of the teacher.}
\label{alg:teacher_opacus}
\KwIn{$D_{\text{priv}}$; classes $K$; sampling rate $q$; steps $S$; clip $C$; noise multiplier $\sigma$; dropout prob.\ $p$; DSAF $(L,\eta_0)$; AW-DP $(\varepsilon_w, w_{\min}, w_{\max})$; lr schedule $\{\eta_t\}$}
\KwOut{DP model parameters $\theta$}
Initialize model $\theta$ and optimizer\;
Create a PyTorch \texttt{DataLoader} for $D_{\text{priv}}$\;
Attach a DP-SGD engine with Poisson subsampling rate $q$, clipping norm $C$, and noise multiplier $\sigma$\;
\For{$t=1$ \KwTo $S$}{
  Sample a minibatch $\mathcal{B}$ using the configured DP sampler (Poisson rate $q$)\;
  Compute minibatch label counts $\{n_k^{(t)}\}_{k=1}^K$ from $\{y_i\}_{i\in\mathcal{B}}$\;
  Set $w_k^{(t)}=\mathrm{clip}\!\left(\frac{|\mathcal{B}|}{K\cdot n_k^{(t)}+\varepsilon_w},\, w_{\min},\, w_{\max}\right)$ (and $w_k^{(t)}=1$ if $n_k^{(t)}=0$)\;
  \ForEach{$(x_i,m_i,y_i)\in\mathcal{B}$}{
    Compute log-Mel $S_i$ and DSAF features $X_i=\mathrm{FixLen}\!\left(\frac{S_i-\mu(S_i)}{\sigma(S_i)+\eta_0},L\right)$\;
    Apply privileged-modality dropout (if privileged input exists) to obtain $m_i'$\;
    Compute per-example loss $\ell_i = w_{y_i}^{(t)}\,\mathrm{CE}\big(y_i, p(X_i,m_i')\big)$\;
  }
  Backpropagate; the DP-SGD engine computes per-sample gradients, clips to $C$, adds Gaussian noise with multiplier $\sigma$, and updates $\theta$\;
}
\Return{$\theta$}\;
\end{algorithm}

\begin{table*}[]
\centering
\caption{Core settings and privacy budgets.}
\label{tab:settings_and_eps}
\footnotesize
\setlength{\tabcolsep}{5pt}
\begin{tabular}{l l @{\hspace{10pt}} c c c}
\toprule
\multicolumn{2}{l}{\textbf{Training / accounting settings}} &
\multicolumn{3}{c}{\textbf{Privacy budgets at epoch 20}} \\
\midrule
$N_{\text{priv}}/N_{\text{aux}}$ & 20k / 5k (recording-disjoint) &
$\sigma$ & Final $\epsilon$ & Strong privacy ($\epsilon\le 1$)? \\
Train batch (expected) & $\mathbb{E}[|\mathcal{B}|]\approx 32$ &
0.5 & 4.9831 & \texttimes \\
Sampling & Poisson, $q=32/N_{\text{priv}}=0.0016$ &
1.0 & 0.4967 & \checkmark \\
Teacher epochs / steps & 20 / 12{,}500 &
3.0 & 0.1209 & \checkmark \\
Clip $C$ & 5.0 &
10.0 & 0.0386 & \checkmark \\
Teacher opt & AdamW, lr $10^{-3}$, wd $10^{-4}$ &
\multicolumn{3}{l}{Non-private baseline: $\sigma=0$ (no DP).} \\
Student KD & $\tau=2$, $\alpha=0.7$ (+ hard-label CE) &
\multicolumn{3}{l}{} \\
Metadata dropout & $p=0.5$ (MM teacher) &
\multicolumn{3}{l}{} \\
\bottomrule
\end{tabular}
\end{table*}

\begin{figure*}[]
  \centering
  \includegraphics[width=0.31\linewidth]{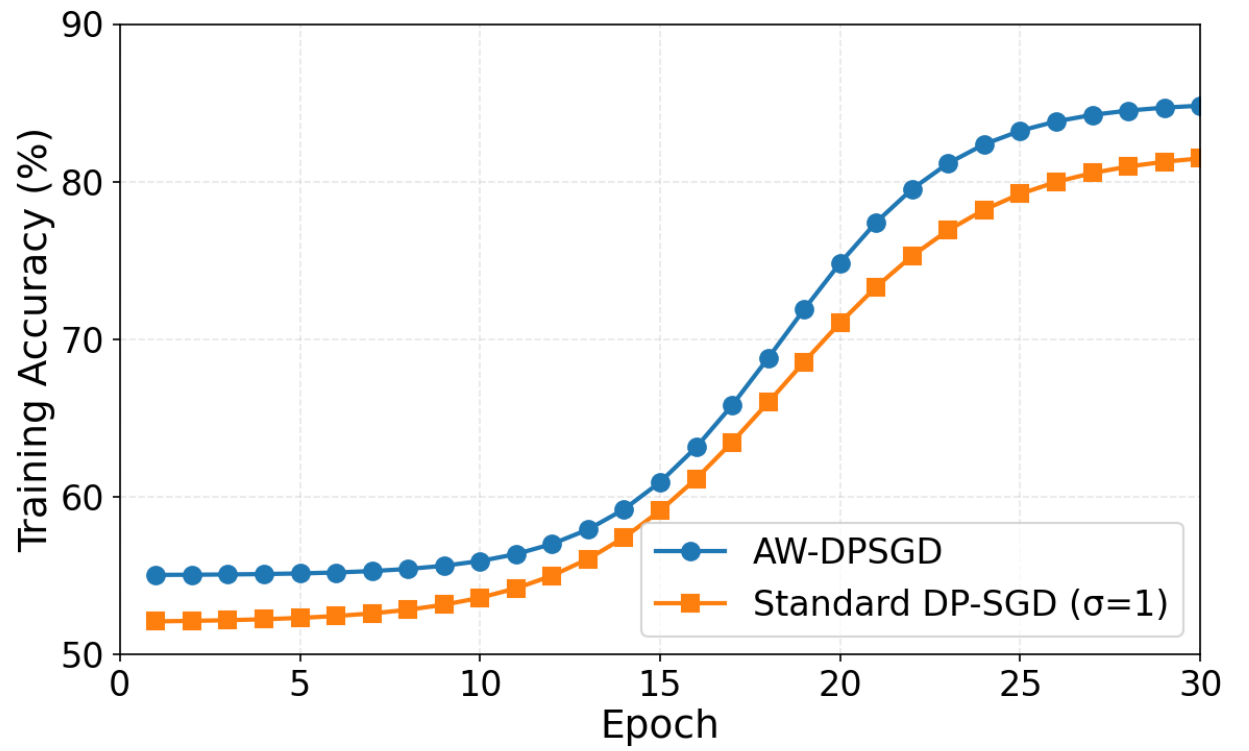}
  \includegraphics[width=0.35\linewidth]{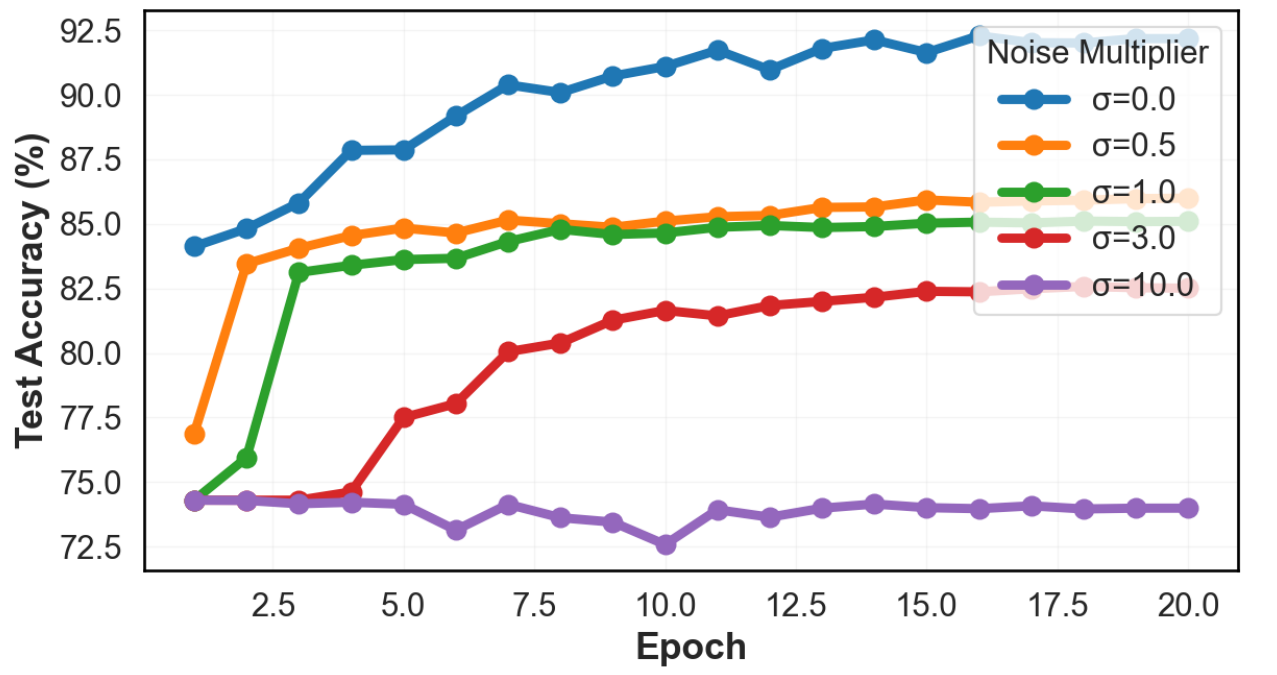}
  \includegraphics[width=0.31\linewidth]{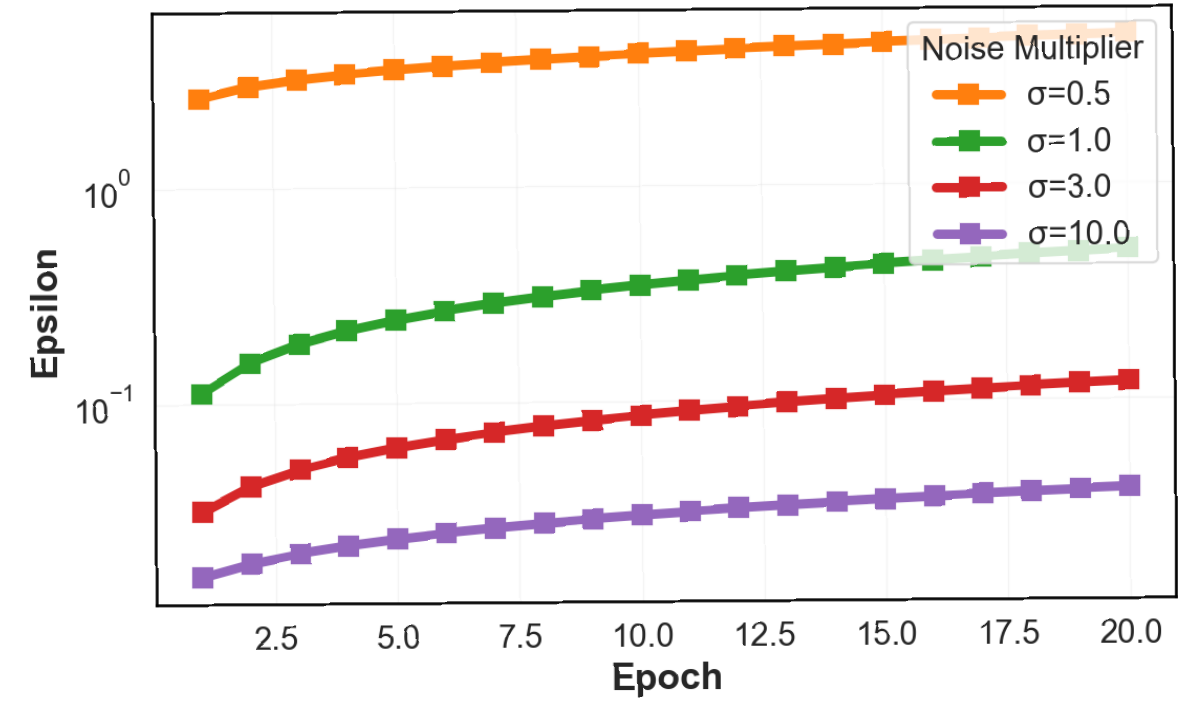}
  \caption{DP teacher (\TA) training dynamics. (left) training accuracy; (middle) test accuracy; (right) composed privacy budget $\epsilon$ vs.\ epoch.}
  \label{fig:curves_combined}
  \vspace{-2mm}
\end{figure*}
\begin{table*}[]
\centering
\caption{Strong privacy results.}
\label{tab:strong_priv_merged}
\footnotesize
\setlength{\tabcolsep}{4pt}
\begin{tabular}{l ccc ccc}
\toprule
\multirow{2}{*}{Setting} &
\multicolumn{3}{c}{$\sigma=1$ ($\epsilon=0.4967$)} &
\multicolumn{3}{c}{$\sigma=3$ ($\epsilon=0.1209$)} \\
\cmidrule(lr){2-4}\cmidrule(lr){5-7}
& Macro-F1$\uparrow$ & Bal-Acc$\uparrow$ & Maj-Pred$\downarrow$
& Macro-F1$\uparrow$ & Bal-Acc$\uparrow$ & Maj-Pred$\downarrow$ \\
\midrule
T-Audio   & $0.3900\pm0.0740$ & $0.3955\pm0.0530$ & $0.9347\pm0.0513$
     & $0.2830\pm0.0000$ & $0.3333\pm0.0000$ & $1.0000\pm0.0000$ \\
S-KD   & $0.4643\pm0.0510$ & $0.4453\pm0.0379$ & $0.8990\pm0.0291$
     & $0.3247\pm0.0098$ & $0.3538\pm0.0050$ & $0.9829\pm0.0043$ \\
T-MM  & $0.4062\pm0.0535$ & $0.4042\pm0.0369$ & $0.9295\pm0.0344$
     & $0.2830\pm0.0000$ & $0.3333\pm0.0000$ & $1.0000\pm0.0000$ \\
S-KD(audio) & $\mathbf{0.4868\pm0.0049}$ & $\mathbf{0.4620\pm0.0085}$ & $\mathbf{0.8799\pm0.0215}$
     & $0.3115\pm0.0342$ & $0.3475\pm0.0177$ & $0.9875\pm0.0150$ \\
S-KD(priv) & $0.4613\pm0.0293$ & $0.4453\pm0.0256$ & $0.8882\pm0.0222$
     & $\mathbf{0.3382\pm0.0586}$ & $\mathbf{0.3623\pm0.0324}$ & $0.9836\pm0.0144$ \\
\bottomrule
\end{tabular}
\end{table*}

\begin{table*}[]
\centering
\caption{Ablation study at two noise levels.}
\label{tab:ablation_merged_bin}
\footnotesize
\setlength{\tabcolsep}{3.5pt}
\begin{tabular}{lccc ccc ccc ccc}
\toprule
& \multicolumn{6}{c}{$\sigma=1.0$ (DP-direct $\epsilon{=}2.973$)} & \multicolumn{6}{c}{$\sigma=3.0$ (DP-direct $\epsilon{=}0.584$)} \\
\cmidrule(lr){2-7}\cmidrule(lr){8-13}
Variant &
\multicolumn{3}{c}{Released (KD)} & \multicolumn{3}{c}{DP-direct} &
\multicolumn{3}{c}{Released (KD)} & \multicolumn{3}{c}{DP-direct} \\
\cmidrule(lr){2-4}\cmidrule(lr){5-7}\cmidrule(lr){8-10}\cmidrule(lr){11-13}
& F1 & BalAcc & MajPred & F1 & BalAcc & MajPred & F1 & BalAcc & MajPred & F1 & BalAcc & MajPred \\
\midrule
B0       & 0.530 & 0.556 & 0.960 & 0.422 & 0.502 & 0.999 & 0.541 & 0.561 & 0.950 & 0.418 & 0.500 & 1.000 \\
DSAF     & 0.592 & 0.587 & 0.845 & 0.418 & 0.500 & 1.000 & 0.571 & 0.573 & 0.881 & 0.418 & 0.500 & 1.000 \\
AW       & 0.599 & 0.597 & 0.912 & 0.429 & 0.505 & 0.997 & 0.652 & 0.637 & 0.870 & 0.418 & 0.500 & 1.000 \\
DSAF\_AW & 0.618 & 0.609 & 0.838 & 0.418 & 0.500 & 1.000 & 0.593 & 0.589 & 0.862 & 0.418 & 0.500 & 1.000 \\
\bottomrule
\end{tabular}
\vspace{-4mm}
\end{table*}

\section{Experiments}
\label{sec:experiments}

\subsection{Experimental Settings}
All experiments are implemented in Python 3.9.6 using PyTorch and Opacus~\cite{yousefpour2021opacus}.

\subsubsection{Datasets and Pre-processing}
We use the Mozilla Common Voice dataset\footnote{\url{http://voice.mozilla.org/}}.
We exclude samples with missing gender labels and treat gender as a 3-way label space (\texttt{male/female/other}).
We extract log-Mel spectrograms with $n_{\text{mels}}{=}40$ Mel bands and standardize their length to $T{=}100$ frames by zero-padding or truncation. Inputs are shaped as $[B,1,40,100]$.

\subsubsection{Private/Auxiliary Split}
We split the training set into a private subset $D_{\text{priv}}$ with $N_{\text{priv}}=20{,}000$ samples and an auxiliary subset $D_{\text{aux}}$ with $N_{\text{aux}}=5{,}000$ samples, using a recording/filename-level disjoint split.
We train the DP teacher on $D_{\text{priv}}$ and train the student non-DP on $D_{\text{aux}}$; we release only the student.

\subsubsection{Privacy Accounting}
We train the teacher with Poisson-subsampled DP-SGD and compute $(\epsilon,\delta)$ using an RDP accountant with $\delta=10^{-5}$.
We use Opacus~\cite{opacus2025} to instrument per-sample gradient computation and DP-SGD updates.
The DP guarantee is scoped to teacher training on $D_{\text{priv}}$ only; the released student incurs no additional privacy loss by post-processing.
We report the exact configuration $(q,S,\sigma,C,\delta)$ in Table~\ref{tab:settings_and_eps}.

\subsubsection{Evaluation Metrics}
We report Macro-F1 and balanced accuracy (Bal-Acc), defined as the unweighted average of per-class recall.
To diagnose collapse, we report Maj-Pred:
\begin{equation}
\mathrm{Maj\text{-}Pred} \;=\; \frac{1}{N}\max_{k\in\{1,\dots,K\}} \#\{i:\hat y_i = k\}.
\end{equation}
In our analysis, we \emph{flag collapse} when $\mathrm{Maj\text{-}Pred}\ge 0.95$ and $\mathrm{Bal\text{-}Acc}$ approaches the degenerate baseline ($\approx 1/K$), indicating a near single-class predictor.

\subsubsection{Compared Methods}
\noindent\textbf{Group A: Single-stage release (DP audio-only model).}
We train an audio-only DP teacher on $D_{\text{priv}}$ and evaluate it directly (\TA).
Unless explicitly disabled in ablations, DP training uses DSAF and AW-DP.

\noindent\textbf{Group B: Two-stage distillation (DP teacher $\rightarrow$ non-DP student; release student only).}
We evaluate:
(1) \TA\ as the DP teacher and its distilled student \SA;
(2) \TMM\ as the DP teacher and its distilled student \SMMa\ (teacher queried with audio only at distillation time);
(3) \SMMp\ (teacher queried with privileged inputs at distillation time; teacher outputs are probabilities only).
Students are trained non-DP on $D_{\text{aux}}$ and are the only released models in Group~B.

\subsection{Results Overview}

\subsubsection{Privacy--Utility Trade-off (Accuracy Context)}
We track training/test accuracy and the composed privacy budget $\epsilon$ over epochs for multiple noise multipliers $\sigma$.

Fig.~\ref{fig:curves_combined} visualizes the privacy--optimization--utility coupling in DP-SGD.
As the noise multiplier $\sigma$ increases, training becomes noisier and learning slows (left), and the resulting test accuracy typically degrades (middle), while the composed privacy budget $\epsilon$ decreases for a fixed training schedule (right).
Importantly, on imbalanced tasks, accuracy alone can obscure a collapse-to-majority solution; therefore we treat accuracy as context only and base utility claims on Macro-F1 and balanced accuracy, together with Maj-Pred as a collapse diagnostic (Table~\ref{tab:strong_priv_merged}).

\subsubsection{Failure Mode under Strong Privacy}
We define the strong-privacy regime as $\epsilon \le 1$.
Under our 3-way setup, $\sigma=1$ yields $\epsilon=0.4967$ and $\sigma=3$ yields $\epsilon=0.1209$ (Table~\ref{tab:settings_and_eps}).
In this regime, training may collapse to a near single-class predictor, where Maj-Pred approaches 1 and class-balanced metrics degrade toward degenerate baselines (e.g., $\mathrm{Bal\text{-}Acc}\rightarrow 1/3$).

\subsubsection{Strong-privacy Results: Distillation Mitigates Collapse}
Table~\ref{tab:strong_priv_merged} compares teacher-only models and distilled students under strong privacy. Distillation reduces collapse (lower Maj-Pred) and improves class-balanced utility relative to single-stage audio-only DP training. Training the DP teacher with privileged metadata can further improve soft-label utility and benefit the released audio-only student.

\subsection{Supplementary: Collapsed binary-label analysis}
\label{sec:supp_binary}
We provide a supplementary binary-label analysis by omitting the rare \texttt{other} class (\texttt{male/female} only). Privacy budgets $\epsilon$ are computed under this binary configuration using Opacus' RDP accountant. \textbf{B0} in Table~\ref{tab:ablation_merged_bin} denotes the base variant (no DSAF, no AW); other rows add the indicated components. Table~\ref{tab:ablation_merged_bin} shows that DP-direct collapses (MajPred $\approx 1$), while offline KD reduces collapse and improves class-balanced utility; DSAF and AW further help.

Sensitivity to auxiliary set size $|D_{\text{aux}}|$. Table~\ref{tab:daux_sensitivity_bin} shows that the released student's performance varies with $|D_{\text{aux}}|$.

\begin{table}[]
\centering
\caption{Sensitivity to $|D_{\text{aux}}|$ for the released student (KD).}
\label{tab:daux_sensitivity_bin}
\footnotesize
\setlength{\tabcolsep}{8pt}
\begin{tabular}{rccc}
\toprule
$|D_{\text{aux}}|$ & KD Macro-F1$\uparrow$ & KD Bal-Acc$\uparrow$ & KD Maj-Pred$\downarrow$ \\
\midrule
1000  & 0.572 & 0.577 & 0.668 \\
5000  & 0.559 & 0.567 & 0.908 \\
10000 & 0.610 & 0.603 & 0.857 \\
\bottomrule
\end{tabular}
\vspace{-2mm}
\end{table}

\section{Conclusion}
\label{sec:conclusion}
By combining a DP teacher trained with DP-SGD and an offline distillation step that releases only an audio-only student, our protocol mitigates strong-privacy majority-class collapse on imbalanced speech while preserving an example-level DP guarantee with respect to $D_{\text{priv}}$ through post-processing. We emphasize that this guarantee does not extend to $D_{\text{aux}}$, which is used for non-private student training.

\bibliographystyle{IEEEbib}
\bibliography{icme2026references}

@inproceedings{dwork2006DP,
  author    = {Cynthia Dwork},
  title     = {Differential Privacy},
  booktitle = {Proceedings of the 33rd International Colloquium on Automata, Languages and Programming (ICALP)},
  series    = {LNCS},
  volume    = {4052},
  pages     = {1--12},
  publisher = {Springer},
  year      = {2006}
}

@inproceedings{dwork2008surveyDP,
  author    = {Cynthia Dwork},
  title     = {Differential Privacy: A Survey of Results},
  booktitle = {Theory and Applications of Models of Computation (TAMC)},
  series    = {LNCS},
  volume    = {4978},
  pages     = {1--19},
  publisher = {Springer},
  year      = {2008}
}

@book{dwork2014,
  author    = {Cynthia Dwork and Aaron Roth},
  title     = {The Algorithmic Foundations of Differential Privacy},
  publisher = {Now Publishers Inc.},
  year      = {2014}
}

@inproceedings{abadi2016deepLearningDP,
  author    = {Mart{\'i}n Abadi and Andy Chu and Ian Goodfellow and H. Brendan McMahan and Ilya Mironov and Kunal Talwar and Li Zhang},
  title     = {Deep Learning with Differential Privacy},
  booktitle = {Proceedings of the 2016 ACM SIGSAC Conference on Computer and Communications Security (CCS)},
  pages     = {308--318},
  publisher = {ACM},
  year      = {2016}
}

@inproceedings{shokri2017membershipInference,
  author    = {Reza Shokri and Marco Stronati and Congzheng Song and Vitaly Shmatikov},
  title     = {Membership Inference Attacks against Machine Learning Models},
  booktitle = {2017 IEEE Symposium on Security and Privacy (SP)},
  pages     = {3--18},
  publisher = {IEEE},
  year      = {2017}
}

@inproceedings{fredrikson2015modelInversion,
  author    = {Matthew Fredrikson and Somesh Jha and Thomas Ristenpart},
  title     = {Model Inversion Attacks that Exploit Confidence Information and Basic Countermeasures},
  booktitle = {Proceedings of the 22nd ACM SIGSAC Conference on Computer and Communications Security (CCS)},
  pages     = {1322--1333},
  publisher = {ACM},
  year      = {2015}
}

@inproceedings{tran2021dpimbalance,
  author    = {Cuong Tran and Minh Dinh and Ferdinando Fioretto},
  title     = {Differentially Private Deep Learning under Skewed Class Distributions},
  booktitle = {Proceedings of the AAAI Conference on Artificial Intelligence (AAAI)},
  volume    = {35},
  number    = {11},
  pages     = {9930--9938},
  year      = {2021}
}

@inproceedings{xia2023perSampleAdaptiveClipping,
  author    = {Tianhao Xia and Shuo Shen and Shuyuan Yao and others},
  title     = {Differentially Private Learning with Per-Sample Adaptive Clipping},
  booktitle = {Proceedings of the AAAI Conference on Artificial Intelligence (AAAI)},
  volume    = {37},
  pages     = {10444--10452},
  year      = {2023}
}

@article{chilukoti2025dpsgdglobadaptv2s,
  author  = {S. V. Chilukoti and M. I. Hossen and L. Shan and others},
  title   = {{DP-SGD-Global-Adapt-V2-S}: Triad Improvements of Privacy, Accuracy and Fairness via Step Decay Noise Multiplier and Step Decay Upper Clipping Threshold},
  journal = {Electronic Commerce Research and Applications},
  volume  = {70},
  pages   = {101476},
  year    = {2025}
}

@inproceedings{chen2020understandingClipping,
  author    = {Xiaojing Chen and Stephen Z. Wu and Mingyi Hong},
  title     = {Understanding Gradient Clipping in Private {SGD}: A Geometric Perspective},
  booktitle = {Advances in Neural Information Processing Systems (NeurIPS)},
  volume    = {33},
  pages     = {13773--13782},
  year      = {2020}
}

@book{OShaughnessy1987SpeechCommunication,
  author    = {Douglas O'Shaughnessy},
  title     = {Speech Communication: Human and Machine},
  publisher = {Addison-Wesley},
  year      = {1987}
}

@inproceedings{ardila2020commonvoice,
  author    = {Rosana Ardila and Megan Branson and Kelly Davis and Michael Henretty and Michael Kohler and Josh Meyer and Reuben Morais and Lindsay Saunders and Francis M. Tyers and Gregor Weber},
  title     = {{Common Voice}: A Massively-Multilingual Speech Corpus},
  booktitle = {Proceedings of the 12th Language Resources and Evaluation Conference (LREC)},
  pages     = {4218--4222},
  year      = {2020}
}

@article{baltrusaitis2019multimodalSurvey,
  author  = {Baltru{\v{s}}aitis, Tadas and Ahuja, Chaitanya and Morency, Louis{-}Philippe},
  title   = {Multimodal Machine Learning: A Survey and Taxonomy},
  journal = {IEEE Transactions on Pattern Analysis and Machine Intelligence},
  volume  = {41},
  number  = {2},
  pages   = {423--443},
  year    = {2019},
  publisher = {IEEE}
}

@article{vapnik2009lupi,
  author  = {Vladimir Vapnik and Akshay Vashist},
  title   = {A New Learning Paradigm: Learning Using Privileged Information},
  journal = {Neural Networks},
  volume  = {22},
  number  = {5--6},
  pages   = {544--557},
  year    = {2009}
}

@misc{hinton2015distilling,
  author       = {Geoffrey Hinton and Oriol Vinyals and Jeff Dean},
  title        = {Distilling the Knowledge in a Neural Network},
  howpublished = {arXiv preprint arXiv:1503.02531},
  year         = {2015},
  url          = {https://arxiv.org/abs/1503.02531}
}

@inproceedings{lopezpaz2016unifying,
  author    = {David Lopez{-}Paz and L{\'e}on Bottou and Bernhard Sch{\"o}lkopf and Vladimir Vapnik},
  title     = {Unifying Distillation and Privileged Information},
  booktitle = {International Conference on Learning Representations (ICLR)},
  year      = {2016}
}

@misc{opacus2025,
  author       = {{Meta Platforms, Inc.}},
  title        = {{Opacus}: Train {PyTorch} models with Differential Privacy},
  howpublished = {\url{https://opacus.ai/}},
  note         = {Accessed: 2025-12-26},
  year         = {2025}
}

@inproceedings{mironov2017rdp,
  title     = {R{\'e}nyi Differential Privacy},
  author    = {Mironov, Ilya},
  booktitle = {2017 IEEE 30th Computer Security Foundations Symposium (CSF)},
  year      = {2017},
  pages     = {263--275}
}

@article{yousefpour2021opacus,
  title   = {Opacus: User-Friendly Differential Privacy Library in PyTorch},
  author  = {Yousefpour, Ashkan and Shilov, Igor and Sablayrolles, Alexandre and Testuggine, Davide and Prasad, Karthik and Malek, Mani and Nguyen, John and Ghosh, Sayan and Bharadwaj, Akash and Zhao, Jessica and Cormode, Graham and Mironov, Ilya},
  journal = {arXiv preprint arXiv:2109.12298},
  year    = {2021}
}

\end{document}